\documentclass[twocolumn,preprintnumbers,amsmath,amssymb,showkeys]{revtex4}
\pdfoutput=1

\usepackage{graphicx}								
\usepackage{dcolumn}								
\usepackage{bm}									

\begin{document}


\title{Periodic Modulation of Extraordinary Optical Transmission through Subwavelength Hole Arrays using Surrounding Bragg Mirrors}

\author{Nathan C. Lindquist, Antoine Lesuffleur and Sang-Hyun Oh}
\email{sang@umn.edu}
\homepage{http://nanobio.umn.edu}
\affiliation{Laboratory of Nanostructures and Biosensing\\Department of Electrical and Computer Engineering\\University of Minnesota, 200 Union St. SE, Minneapolis, MN 55455, U.S.A.}

\date{\today}

\begin{abstract}
The enhanced light transmission through an array of subwavelength holes surrounded by Bragg mirrors is studied, showing that the mirrors act to confine the surface plasmons associated with the Extraordinary Optical Transmission effect, forming a surface resonant cavity. The overall effect is increased light transmission intensity by more than a factor of three beyond the already enhanced transmission, independent of whether the Bragg mirrors are on the input or the output side of the incident light. The geometry of the Bragg mirror structures controls the enhancement, and can even reduce the transmission in half. By varying these geometric parameters, we were able to periodically modulate the transmission of light for specific wavelengths, consistent with the propagation and interference of surface plasmon waves in a resonant cavity. FDTD simulations and a wave propagation model verify this effect.
\end{abstract}



\maketitle

From its initial discovery by Ebbesen et al. \cite{ebbesen}, the Extraordinary Optical Transmission (EOT) effect has generated a lot of interest and research, both for its potential application in photonic devices, and also for understanding its underlying physical mechanism. This effect manifests itself as increased light transmission at specific wavelengths through a periodic array of subwavelength holes in a thin metal film, which is significantly larger than that predicted by conventional Bethe aperture theory \cite{bethe}. It is known that geometric factors play a critical role in this effect, such as the periodicity of the hole array \cite{ebbesen}, the film thickness \cite{martinmoreno}, the presence of bumps and dimples on the metal surface \cite{grupp1}, and the shape and orientation of the holes \cite{koerkamp,gordon,degiron}. It is generally accepted that the EOT effect is mediated by surface plasmons (SPs) generated at the metal-dielectric interface by the periodic array of nanoholes \cite{ghaemi,gao,grupp,barnes} although there are other theories that explain the enhanced transmission in both holes and slits without the involvement of SPs \cite{lezec} or with their negative role \cite{cao}. It is also known that SPs can be manipulated like any other propagating wave: they can reflect off structures that act as Bragg mirrors \cite{sanchezgil}, be confined by wall-like structures \cite{huang}, can interfere to form standing waves \cite{weeber}, and can even be confined in nanocavities \cite{hideki}. Combining the various properties and the unique effects associated with SP waves can lead to a deeper understanding of the basic physics involved and is important for the development of new plasmonic devices, such as label-free biosensors \cite{lesuffleur} and ultimately, nanophotonic circuitry.

In this letter we study the transmission of light through an array of nanoholes surrounded by Bragg mirrors and report the realization of a lateral SP resonant cavity in combination with the EOT effect. The Bragg mirrors provide a mechanism to confine the SPs coupled to the array of nanoholes and prevent their energy escaping from the area of the nanoholes. The overall effect is increased light transmission by more than a factor of three beyond the already enhanced EOT effect. The light transmission is found to depend strongly on the geometry of the Bragg mirror lateral resonant cavity structure. By varying the geometric parameters, we are able to periodically modulate the transmission of light for specific resonant wavelengths, and search for optimal ÒtuningÓ conditions. Finite difference time domain (FDTD) simulations confirm the confinement of SP waves and the modulation effect.

\begin{figure*}[H]
\includegraphics[width=0.5\textwidth]{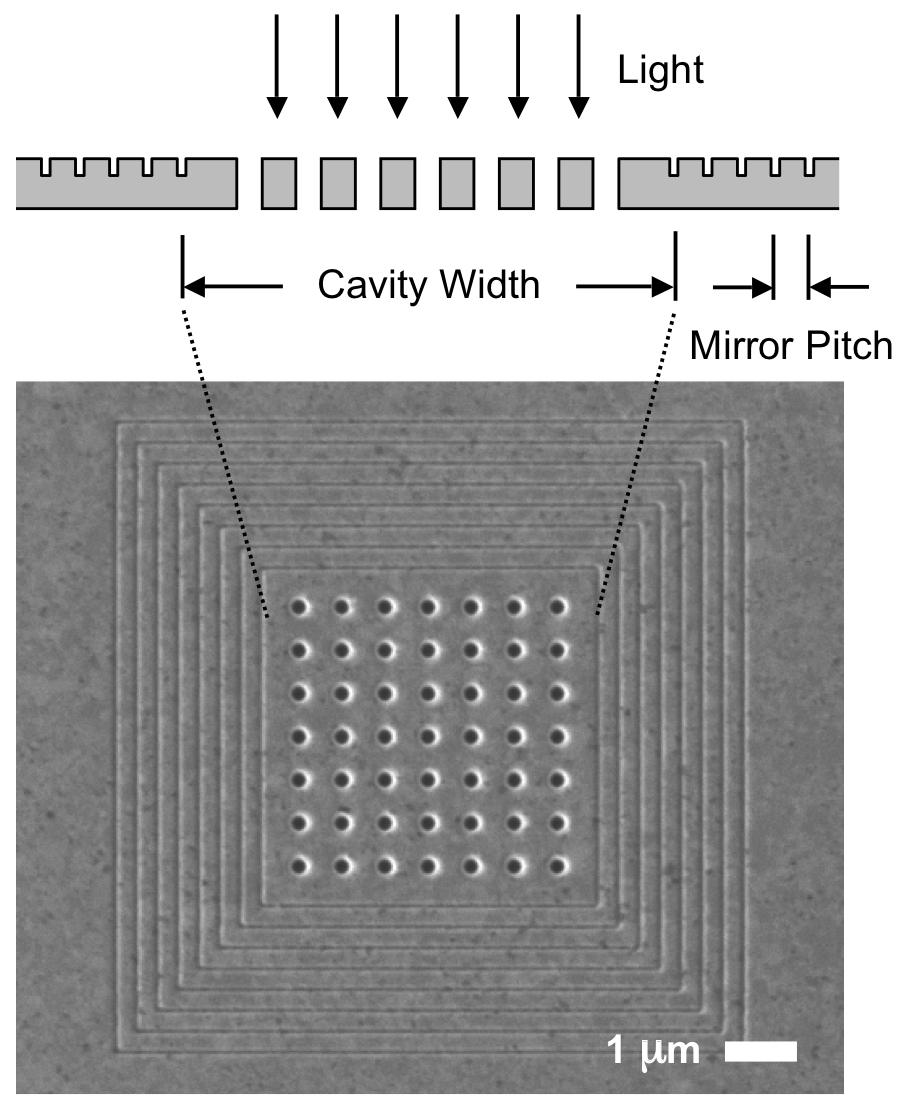}
\caption{\label{figure1} A scanning ion beam image of the nanohole array surrounded by narrow Bragg mirror grooves on a 100 nm thick gold film on a glass substrate. The nanohole array has a periodicity of 635 nm, and the holes have a diameter of 200 nm. The periodic nanohole array couples incident light into SPs, and the surrounding Bragg mirrors, with a pitch of 300 nm and width of 60 nm, confine the SPs within the nanohole array, enhancing light transmission. The diagram shows a two-dimensional profile, an important geometric parameter being the cavity width. By adjusting this distance, it is possible to periodically modulate the extraordinary optical transmission through the nanohole array.}
\end{figure*}

Figure \ref{figure1} shows a scanning ion beam picture of a nanohole array surrounded by Bragg mirrors. It should be noted that although this design looks similar in form to the so-called ``Bull's Eye'' structure \cite{lezec2}, it is different in function. In the Bull's Eye structure design, the periodic grooves are used to either generate SPs leading to enhanced optical transmission or to enhance directionality upon re-radiation through a subwavelength aperture  \cite{lezec2,garciavidal}, depending on whether the grooves are on the input (illuminated) side or the output side. In our design, the grooves, due to their specific periodicity, are used as Bragg mirrors that reflect and confine the SP waves generated by the nanoholes themselves, leading to enhanced transmission independent of which side is illuminated, as is demonstrated below.

The samples were created with focused ion beam milling on a 100 nm thick gold film with a 5 nm Cr adhesion layer on a glass substrate. Each nanohole array was made with 200 nm diameter circular holes with a 635 nm square periodicity. According to standard calculations, the wavelength of a SP, $\lambda_{SP}$, coupled to a grating at normal incidence is equal to the period of the grating, which, for the (1,0) resonance, is simply the periodicity of the nanohole array \cite{ebbesen,thio}. The size of the array, i.e. the number of holes, was chosen with a trade-off in mind: if the number of holes was too large, the SP waves wouldn't propagate far compared to the size of the array, due to high losses, and the effect of the mirrors would be small; however, if the array was too small, incident light wouldn't couple to produce SPs efficiently, and the EOT effect and its transmission peaks wouldn't be as pronounced \cite{lezec}. The final array size of 7-by-7, in our experiments, seemed to balance this tradeoff, erring on the side of keeping the array as small as possible. This array size effect is discussed further below, along with the presented data.

The Bragg mirror grooves each have a width of 60 nm, and are milled halfway through the gold film. We ensured that no detectable light came through these grooves by characterizing test samples milled only with grooves. According to the discussion in \cite{lopeztejeira}, a periodic array of grooves can optimally reflect SPs when:
\begin{equation}
	k_{SP} D = m \pi,
	\label{equation1}
\end{equation}
where $k_{SP} = 2 \pi / \lambda_{SP}$ is the wavevector of the SP, $D$ is the periodicity of the grooves, and $m = 1, 2, 3, \dots$ In our case, $m = 1$, and the periodicity of the grooves is $\lambda_{SP}/2$, which maximizes reflectivity \cite{weeber}. The grooves are placed to form eight concentric squares around the 7-by-7 nanohole array, which creates a two-dimensional confinement structure on the surface of the gold film and defines the cavity width as the width of the inner square. The samples were illuminated using a tungsten-halogen lamp and a microscope objective (50x, NA = 0.55). The transmitted light was collected using an optical fiber (200-$\mu$m diameter core), and the zero-order transmission spectrum was analyzed with an Ocean Optics USB Fiber Optic Spectrometer.

\begin{figure*}[H]
\includegraphics[width=0.5\textwidth]{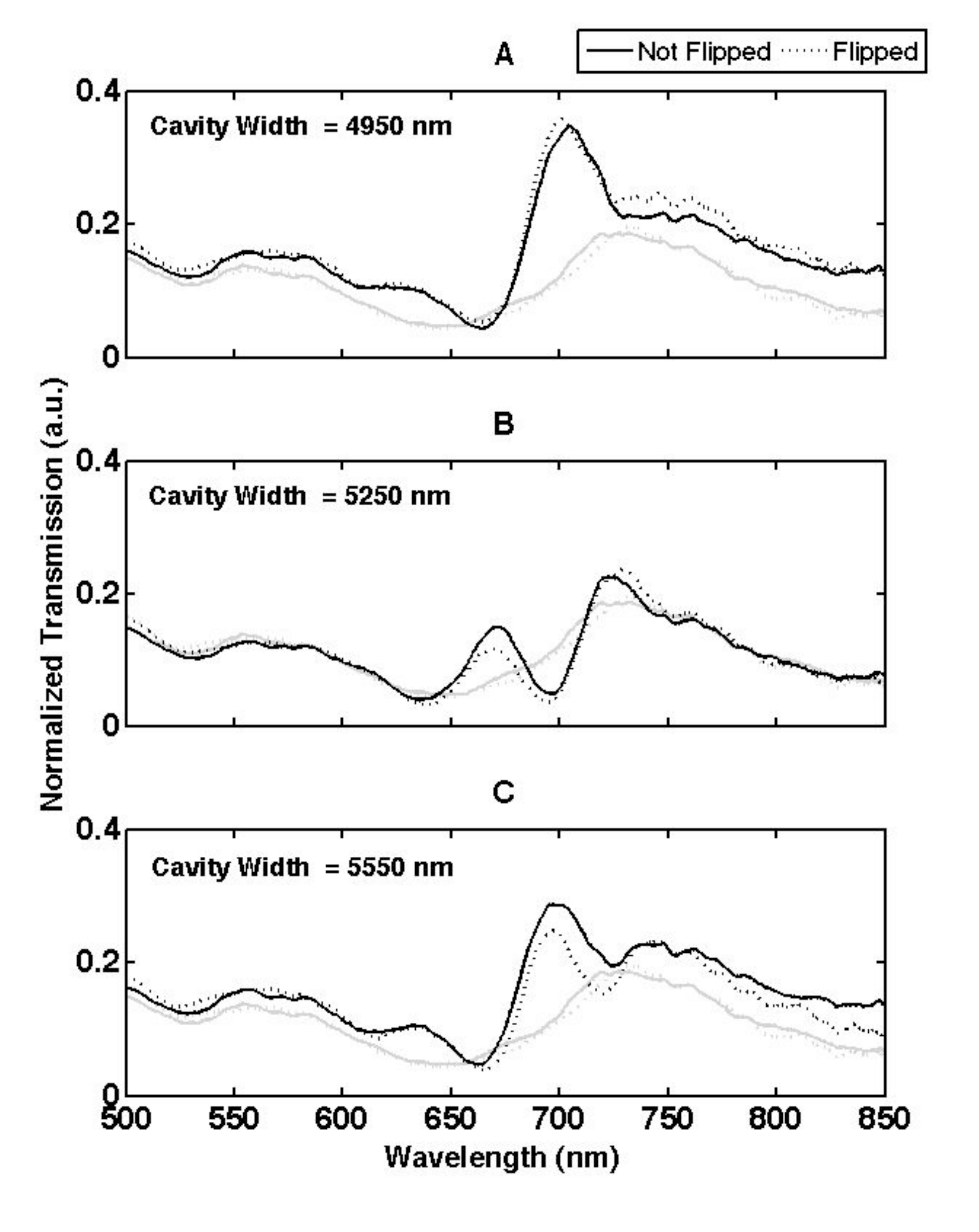}
\caption{\label{figure2} Plots of the normalized transmitted spectra for nanohole arrays both with surrounding Bragg mirrors (black lines) and without (grey lines). The plots also include an un-flipped sample, where the grooves are directly illuminated, and a flipped sample, where the grooves are not directly illuminated. The 7-by-7 nanohole array without surrounding Bragg mirrors shows two peaks, corresponding to the (1,1) and (1,0) EOT transmission maxima. The transmitted spectra with Bragg mirrors depend strongly on the cavity width as it is varied from 4450 nm to 5650 nm. Graphs A, B and C show the progression through one period of modulation, with A and C corresponding to a transmission maximum at 700 nm, and B corresponding to a transmission minimum. One period of modulation is completed when the cavity width changes by 635 nm. This modulation, near the (1,0) EOT peak at 700 nm, is independent of which side, top or bottom, of the device is illuminated.}
\end{figure*}

Figure \ref{figure2} shows the spectra (normalized to the spectrum of the lamp) for a 7-by-7 nanohole array both with and without Bragg mirrors for three cavity widths (4950 nm, 5250 nm, and 5550 nm). The 7-by-7 nanohole array without surrounding Bragg mirrors shows two peaks near 550 nm and 730 nm, which correspond to the (1,1) and the (1,0) transmission resonances of the array \cite{ebbesen, koerkamp, gordon, chang}. These resonances are shifted to longer wavelengths than expected theoretically due to the small number of holes per array \cite{lezec}. Spectra from both a 7-by-7 and a 16-by-16 hole array (not shown here) confirmed this dependency, and also show that increasing the number of holes sharpens the standard EOT resonance peaks. The spectra in figure \ref{figure2} show that the cavity width has a distinct effect on the transmission peaks. Figures \ref{figure2}A and \ref{figure2}C show enhanced transmission near 700 nm, whereas figure \ref{figure2}B shows suppressed transmission. Figures \ref{figure2}A and \ref{figure2}C also show that including the surrounding Bragg mirrors, the (1,0) resonance peak shifted to a shorter wavelength (700 nm), and was sharpened considerably. The spectra presented in figure \ref{figure2} suggest that the mirrors may contribute, effectively, to increase the number of holes in the smaller 7-by-7 array.

Data for a ÒflippedÓ sample is also shown on figure \ref{figure2}, in which the grooves themselves are not directly illuminated. The spectra for the flipped and un-flipped configurations are nearly identical (especially figure \ref{figure2}A) in their important features, showing the same modulation effect, and the same spectral shape. There is a slight intensity variation across a few areas of the spectra, but this is even seen with the 7-by-7 reference sample. Therefore the operation of our device is different than illuminating periodic grooves to generate SPs and enhance transmission, where the transmission spectra change dramatically depending on whether or not the grooves are illuminated \cite{garciavidal}. This hints that an even more pronounced effect could be seen by the introduction of Bragg mirrors on both sides of the film.

\begin{figure*}[H]
\includegraphics[width=0.6\textwidth]{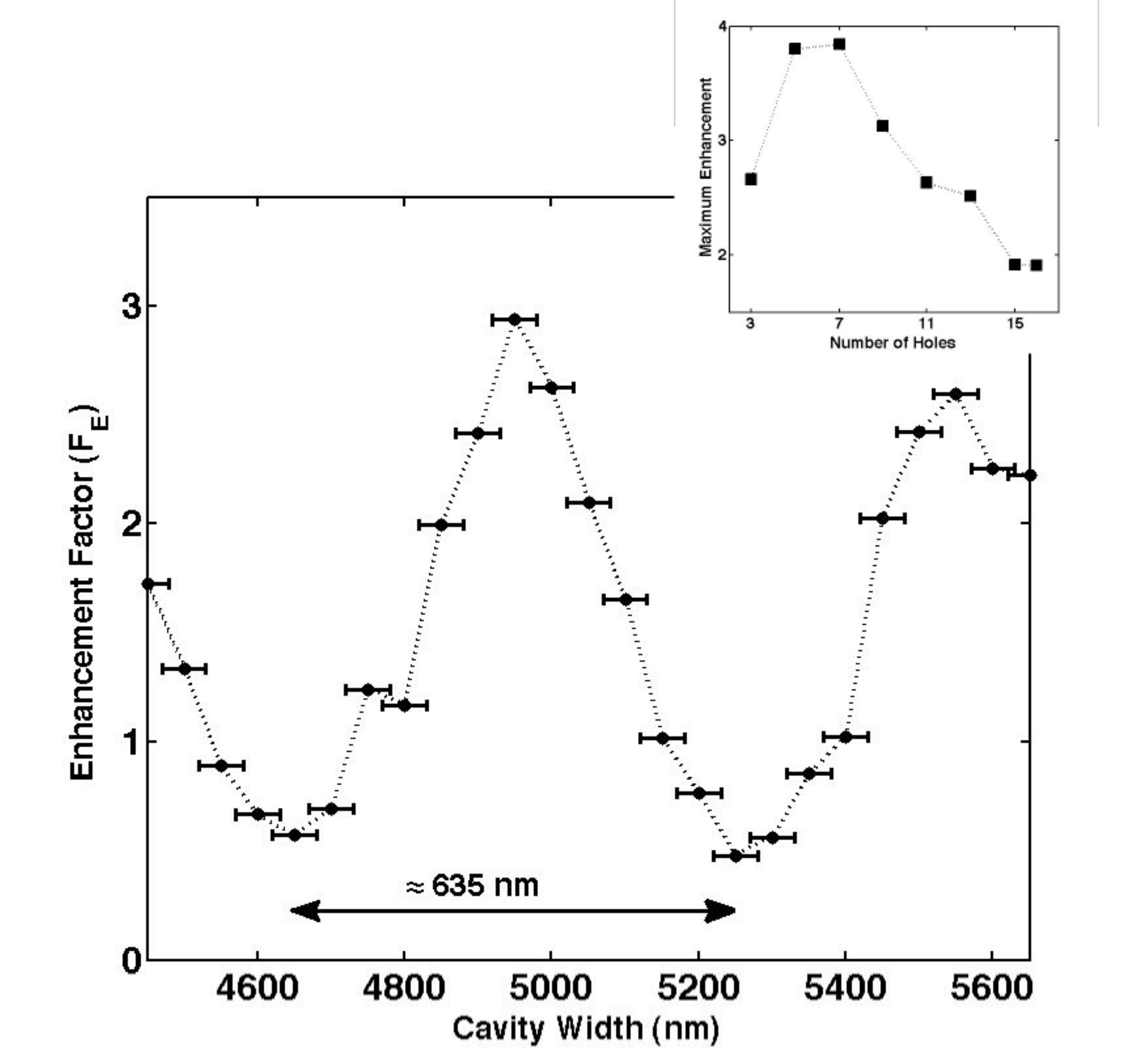}
\caption{\label{figure3} Plot of the enhancement factor $F_E$ (transmission with Bragg mirrors normalized to transmission without) of the (1,0) EOT peak versus cavity width. Tracking $F_E$ shows a periodic modulation, with a period of 635 nm, as the cavity width is varied from 4450 nm to 5650 nm. The data points correspond to individual samples for each 50 nm step. The maximum enhancement factor is $F_E$ = 3.0, whereas the minimum corresponds to a reduction in transmission with $F_E$ = 0.5. (inset): Plot showing the influence of the number of holes on the resonant cavity effect, with the optimal number showing $F_E$ nearly equal to 4.}
\end{figure*}

Figure \ref{figure3} shows the enhancement factor $F_E$, defined as transmission of an array with surrounding Bragg mirrors normalized to an array without, of the (1,0) resonance peak versus the cavity width, every 50 nm from 4450 nm to 5650 nm. Each data point corresponds to a different, individually milled sample, showing a consistent effect. As the width of the cavity is varied, the mirror-enhanced peak goes through a series of maxima and minima. The maximum (constructive) effect corresponds to an enhancement of $F_E = 3.0$, and the minimum (destructive) effect corresponds to $F_E = 0.5$. The period of modulation is $\approx$635 nm, which matches the periodicity of the nanohole array and the wavelength of the SP that mediates the extraordinary transmission of light at the (1,0) resonance. The trend of figure \ref{figure3} is reminiscent of a transmission curve for a Fabry-Perot cavity with losses. In our case, however, the nanohole array is sampling the field {\em inside} the cavity, and coupling to it through itÕs {\em own} (1,0) resonance, creating an intriguing combination of the two effects. This $\approx$635 nm periodic modulation of an optical signal with a wavelength of 700 nm points towards an interference phenomenon between SP waves propagating inside a resonant cavity defined by the surrounding Bragg mirrors. Indeed, all geometric parameters were tuned specifically for a surface wave of wavelength 635 nm.

To explain the modulation observed in figure \ref{figure3}, we consider the phase difference between a SP reflected by the Bragg mirror compared to the phase of a SP launched from the nanohole array. As discussed in \cite{chang}, for a 200 nm diameter hole and visible wavelengths, the edge of the circular hole can be considered as the source of the SPs. The total phase picked up by a SP propagating from the edge of a circular hole, reflecting once from the Bragg mirror, and propagating back can be written as:
\begin{equation}
	\phi = m \pi + 2 \pi (2 L / \lambda_{SP} )
	\label{equation2}
\end{equation}
where $L$ is the distance from the edge of the hole to the Bragg mirror, and $m \pi$, the same $m$ as in equation (\ref{equation1}), is picked up upon reflection \cite{lopeztejeira}. The distance $L$ depends directly on the cavity width since all the other geometric parameters are fixed (the periodicity of the array and the hole diameter). The condition for constructive interference at the nanohole and the resonant cavity condition is therefore $\phi = 2 n \pi$, where $n = 1, 2 , \dots$ which rearranges to $L_{constructive} = \lambda_{SP} (2 n - 1)/4$. Taking the periodicity of the 7-by-7 array and the hole diameter as 635 nm and 200 nm, respectively, gives the condition for cavity resonance: Cavity Width = $\lambda_{SP} (n - 1/2) + 4010$ nm. For $n = 2$, and $\lambda_{SP} = 635$ nm, the calculated value of 4960 nm matches very well with the experimental value of 4950 $\pm$ 30 nm. This indicates that even such a simple equation as (\ref{equation2}), which only considers a phase model for the propagating SP waves, can lead to physical insight and aid in the design of the SP resonant cavity.

Since the Bragg mirrors reflect the SP leaving the edges of the nanohole array, the number of holes in the array is an important parameter. Indeed Bravo-Abad and co-workers have shown that the edges and the finite size of the array can have a profound influence on the spatial distribution of the transmitted light \cite{bravoabad}. Figure \ref{figure3} (inset) presents the maximum enhancement factor measured for various Bragg resonators wherein the number of holes was changed. New samples were fabricated, taking into account the insight of equation (\ref{equation2}), leading to a maximum enhancement of nearly 4 for the 7-by-7 array. The main trend of this curve confirms the trade-off considered previously, namely that with either too few or too many holes, the Bragg mirrors wouldn't be able to efficiently create a resonant cavity. There are three main areas of the graph: (1) for less than 5 holes per array, the EOT mechanism doesn't couple efficiently enough to the array to produce SPs which would resonate in the cavity; (2) for an array sizes between 5-by-5 and 7-by-7, the EOT SPs are coupled efficiently to the array and into the resonant cavity, and the overall light transmission is enhanced significantly; (3) For more than 9-by-9 holes, the SPs propagating in the array suffer too many losses, and the Bragg mirrors, instead of creating a resonant cavity, merely affect the edges of the array, with only moderately enhanced transmission.

\begin{figure*}[H]
\includegraphics[width=0.8\textwidth]{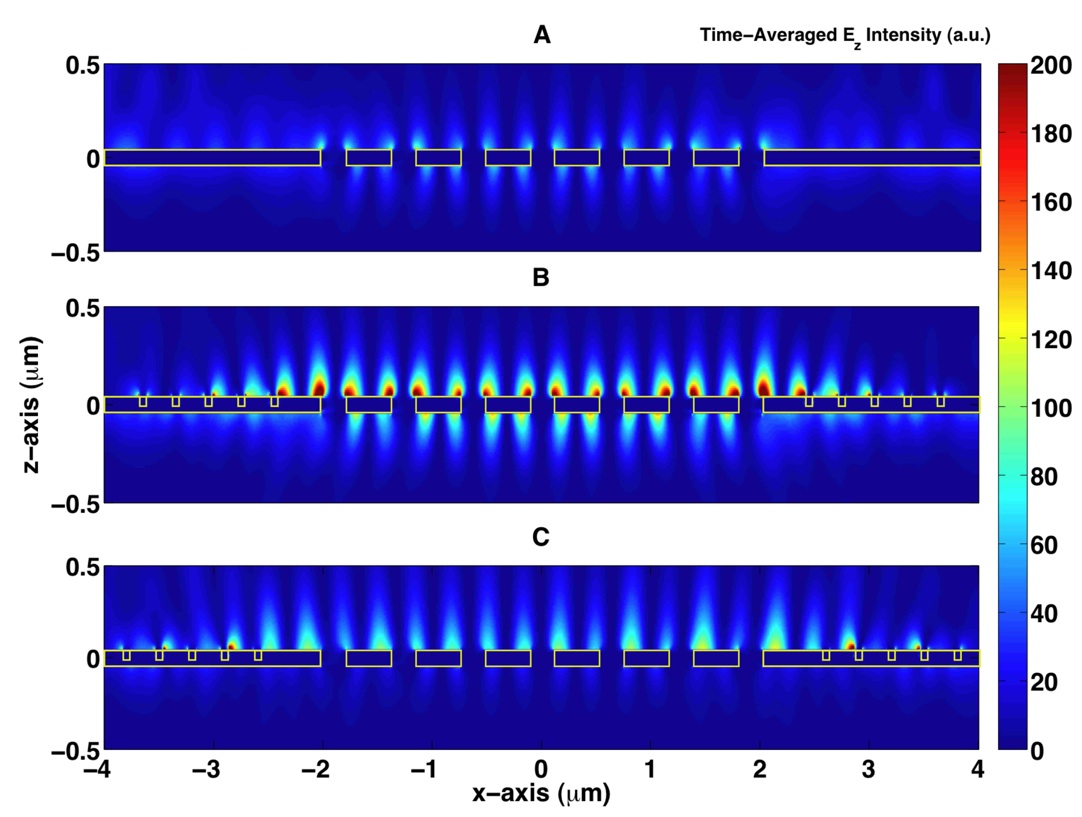}
\caption{\label{figure4} FDTD simulation results showing the time-averaged intensity of the evanescent surface plasmon (SP) field for three configurations: (A) a bare nanohole array, (B) a maximally resonant SP cavity (cavity width = 4900 nm), and (C) a SP cavity with destructive interference (cavity width = 5200 nm). All plots are shown with the same intensity scaling for comparison, demonstrating that the SPs are well confined in a lateral resonant cavity across the entire array and either enhanced or suppressed by the surrounding Bragg mirrors. The intensity on the lower output side of the film is the highest for the constructive interference effect, showing increased transmission.}
\end{figure*}

Figure \ref{figure4} shows FDTD simulations of the effects of the Bragg mirrors both at the edges of the array, and for the overall modulated transmission. Light, with wavelength 680 nm, is incident from the top (+$z$), illuminating both the Bragg grooves and the 635 nm periodicity nanoholes. The grid size for calculations is 10 nm in $x$ and $z$, and 20 nm in $y$. Periodic boundary conditions are used in the $y$ direction, and there are 7 holes in the $x$ direction, confined laterally by the Bragg mirrors. The three plots, (A) a plain nanohole array, (B) a constructive cavity, width 4900 nm, and (C) a destructive cavity, width 5200 nm, show an $x$-$z$ slice through the gold film of the time-averaged intensity of the $z$-component of the electric field all plotted with the same intensity scaling. The SP fields, compared to that of the bare array, become more intense and localized for the constructive case, and less intense and ``broadened'' for the destructive case. The Bragg mirrors are seen to confine the SPs that would have otherwise escaped from the array.  On the lower, output side, the highest field intensity occurs for the constructive cavity (B), thereby transmitting the most light.

\begin{figure*}[H]
\includegraphics[width=0.5\textwidth]{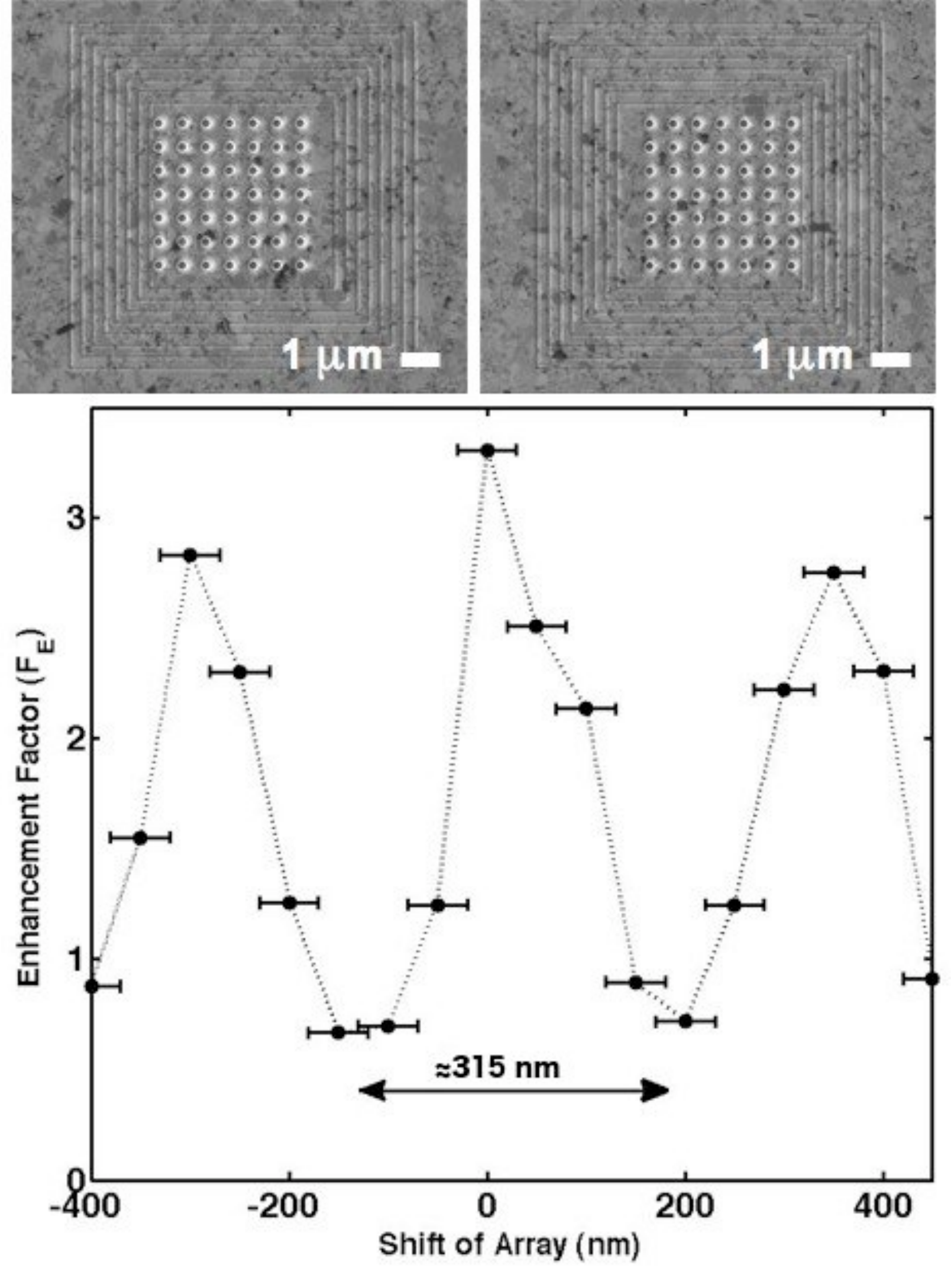}
\caption{\label{figure5} Plot of the enhancement factor $F_E$ of the (1,0) EOT peak versus position of the nanohole array within the cavity. By shifting the nanohole array within the surrounding Bragg mirrors, $F_E$ is again modulated, this time completing one period every 315 nm. The scanning ion beam images show two different samples where the nanohole array is shifted from one side of the surrounding Bragg mirrors to the other by $\pm$300 nm. The maximum enhancement factor seen here is $F_E = 3.2$, whereas the minimum is again a reduction in transmission with $F_E = 0.75.$ }
\end{figure*}

Further confirmation that interfering SPs cause the modulation and cavity effect came from a series of different samples wherein the position of the nanohole array was shifted from the center of the cavity by -400 nm to +450 nm along one direction. Figure \ref{figure5} shows two such samples and the enhancement factor $F_E$ for each consecutive shift of 50 nm within the cavity. Again, a periodic modulation of the transmission peak corresponding to the (1,0) EOT resonance is observed, this time with a period of $\lambda_{SP}/2 = 315$ nm, which is consistent with the prediction of equation (\ref{equation2}). The largest enhancement is $F_E = 3.3$, where the minimum is $F_E = 0.7$. Notably, by using polarized light to generate SPs traveling perpendicular to the shift of the nanohole array, there was no modulation.

In conclusion, we present a novel device that enables the periodic modulation of the standard EOT effect by the introduction of surrounding Bragg mirrors that act as a lateral resonant cavity for SPs generated by the array of nanoholes. Outward propagating SPs reflect back into the array and interfere either constructively or destructively, depending on the geometry of the mirror structure, leading to enhanced or reduced transmission at certain wavelengths. We discussed the underlying physics and key geometric parameters most important to this periodic modulation effect. Also, we have demonstrated experimentally that no SPs are launched by the grooves for generating this effect, which make this device significantly different than other structures combining periodic corrugations and one aperture. The ability to confine the SP energy within a nanoscale device and periodically modulate EOT using an integrated lateral resonant cavity may lead to new concepts and designs that harness the effect of SPs in nanophotonic devices and circuitry.

\begin{acknowledgments}
We would like to thank Prof. Anand Gopinath for his insightful discussions and allowing access to the FDTD simulation tools.  Device fabrication was performed at the NanoFabrication Center at the University of Minnesota (a member of the National Nanotechnology Infrastructure Network), which is supported by the National Science Foundation.
\end{acknowledgments}

\newpage

\newpage

\newpage

\newpage

\newpage

\newpage

\bibliography{paper}	

\end{document}